\documentclass[prd,aps,floats,preprint,preprintnumbers]{revtex4}
\usepackage{epsfig}
\usepackage{amsmath} 
\usepackage{amssymb}
\newcommand{\be}{\begin{equation}}
\newcommand{\ee}{\end{equation}}
\newcommand{\bea}{\begin{eqnarray}}
\newcommand{\eea}{\end{eqnarray}}

\newcommand{\mpl}{M_{\rm P}}

\begin{document}

\setlength{\unitlength}{1mm}

\title{The Cosmology of Generalized Modified Gravity Models}

\author{Sean M. Carroll$^1$\footnote{carroll@theory.uchicago.edu}, 
Antonio De Felice$^2$\footnote{defelice@physics.syr.edu}, Vikram Duvvuri$^1$\footnote{duvvuri@theory.uchicago.edu}, \\
Damien A. Easson$^2$\footnote{easson@physics.syr.edu}, Mark Trodden$^2$\footnote{trodden@physics.syr.edu} 
and Michael S. Turner$^{1,3,4}$\footnote{mturner@oddjob.uchicago.edu}} 

\affiliation{$^1$Enrico Fermi Institute, Department of Physics, 
and Kavli Institute for Cosmological Physics, University of Chicago, 
5640 S. Ellis Avenue, Chicago, IL 60637-1433, USA. \\ 
$^2$Department of Physics, Syracuse University, 
Syracuse, NY 13244-1130, USA. \\ 
$^3$Department of Astronomy \& Astrophysics, University of Chicago, 
Chicago, IL 60637-1433, USA. \\ 
$^4$NASA/Fermilab Astrophysics Center, 
Fermi National Accelerator Laboratory, Batavia, IL 60510-0500, USA.}

\begin{abstract} 
We consider general curvature-invariant modifications of the Einstein-Hilbert action 
that become important only in regions of extremely low space-time curvature. We investigate the
far future evolution of the universe in such models, examining the possibilities for cosmic acceleration
and other ultimate destinies. The models generically possess de Sitter space as an unstable solution
and exhibit an interesting set of attractor solutions which, in some cases, provide alternatives to dark 
energy models.
\end{abstract}

%\pacs{PACS Numbers: }

\maketitle

\section{Introduction}
The acceleration of the universe presents one of the greatest problems in theoretical physics today. The increasingly accurate observations of type Ia supernovae light-curves, coupled with exquisite measurements of CMB anisotropies and large scale structure data~\cite{Riess:1998cb,Perlmutter:1998np,Tonry:2003zg,Bennett:2003bz,Netterfield:2001yq,Halverson:2001yy} have forced this issue to the forefront of those facing particle physicists, cosmologists and gravitational physicists alike.

This problem has been attacked head on, but no compelling, well-developed and well-motivated solutions have yet emerged. While much work has focused on the search for new matter sources that yield accelerating solutions to general
relativity, more recently some authors have turned to the complementary approach of examining whether new gravitational physics might be responsible for cosmic acceleration.

There have been a number of different 
attempts~\cite{Deffayet:2001pu,Freese:2002sq,Arkani-Hamed:2002fu,Dvali:2003rk,Nojiri:2003wx,Carroll:2003wy,Capozziello:2003tk,Arkani-Hamed:2003uy,Abdalla:2004sw}
to modify gravity to yield accelerating cosmologies at late times. The path we are concerned with in this paper is the direct addition of 
higher order curvature invariants to the Einstein-Hilbert action. The first example of this was provided by the model of Carroll, Duvvuri, 
Trodden, and Turner (CDTT)~\cite{Carroll:2003wy} (see also~\cite{Capozziello:2003tk}). For subsequent work on various aspects and 
extensions of this model, see 
\cite{Vollick,Dick,Nojiri_1,Dolgov,Nojiri_2,Meng1,Chiba,Meng2,Flanagan,Woodard,Ezawa,Flanagan_2,Rajaraman,Vollick_2,Nunez,Allemandi,Lue:2003ky}. 
In particular, the simplest model has been shown to conflict with solar system tests of gravity~\cite{Dolgov,Chiba,Woodard}.  Our approach is purely 
phenomenological. The evidence for cosmic acceleration is very sound. In pure Einstein gravity, as matter dilutes away in the expanding universe, 
the expansion rate inevitably slows. Our question is can this be avoided within the gravitational sector of the theory, as opposed to adding new 
energy sources. One way is by adding a cosmological constant. In this paper we explore a wider class of modifications that share the feature of late-time 
accelerating behavior, thus fitting cosmological observations.

Consider a simple correction to the Einstein-Hilbert action,
\begin{equation}
\label{action} S =\frac{\mpl^2}{2}\int d^4 x\,
\sqrt{-g}\left(R-\frac{\mu^4}{R}\right) +\int d^4 x\, \sqrt{-g}\,
{\cal L}_M \ . 
\end{equation} 
where $\mu$ is a new parameter with units of $[{\rm mass}]$ and ${\cal L}_M$ is the Lagrangian density for matter. This action gives rise to fourth-order equations of motion. In the case of an action depending purely on the Ricci scalar (and on its derivatives) it is possible to transform from the frame used in~(\ref{action}), which we call the {\em matter frame}, to an
{\em Einstein frame}, in which the gravitational Lagrangian takes the
Einstein-Hilbert form and the additional degrees of freedom ($\ddot H$
and $\dot H$) are represented by a scalar field $\phi$.
The details of this can be found in~\cite{Carroll:2003wy}. The scalar field is  minimally coupled to
Einstein gravity, non-minimally coupled to matter, and has a potential given by
\begin{equation} 
\label{potential} 
V(\phi)=\mu^2 \mpl^2
\exp\left(-2\sqrt{\frac{2}{3}}\frac{\phi}{\mpl} \right)\sqrt{\exp
\left(\sqrt{\frac{2}{3}}\frac{\phi}{\mpl} \right)-1} \ .
\end{equation} 

Consider vacuum cosmological solutions. 
We must specify the initial values of $\phi$ and $\phi'$, denoted
as $\phi_i$ and ${\phi'}_i$.  For simplicity we take $\phi_i \ll
\mpl$. There are three qualitatively distinct outcomes, depending
on the value of ${\phi'}_i$.

{\em 1.  Eternal de Sitter.}  There is a critical value of ${\phi'}_i
\equiv {\phi'}_C$ for which $\phi$ just reaches the maximum of the
potential $V(\phi)$ and comes to rest.  In this case the Universe
asymptotically evolves to a de~Sitter solution. This
solution requires tuning and is unstable, since any perturbation will
induce the field to roll away from the maximum of its potential.

{\em 2.  Power-Law Acceleration.}  For ${\phi'}_i > {\phi'}_C$, the
field overshoots the maximum of $V(\phi )$ and the Universe evolves to
late-time power-law inflation, with observational consequences similar
to dark energy with equation-of-state parameter $w_{\rm DE}=-2/3$.

{\em 3.  Future Singularity.}  For ${\phi'}_i < {\phi'}_C$, $\phi$
does not reach the maximum of its potential and rolls back down to
$\phi =0$.  This yields a future curvature singularity.

In the more interesting case in which the Universe
contains matter, it is possible to show that the three
possible cosmic futures identified in the vacuum case remain in the
presence of matter. 

By choosing $\mu\sim 10^{-33}\,$eV, the corrections to the standard
cosmology only become important at the present epoch, making this
theory a candidate to explain the observed acceleration of the
Universe without recourse to dark energy.  Since we have no particular
reason for choosing this value of $\mu$, such a tuning is certainly not attractive.
However, it is worth commenting that this small correction to the action, 
unlike most small corrections in physics, is destined to be important as 
the Universe evolves.

Clearly our choice of correction to the gravitational action can be
generalized.  Terms of the form $-\mu^{2(n+1)}/R^n$, with $n>1$, lead
to similar late-time self acceleration, with behavior similar to a dark energy 
component with equation of
state parameter 
\begin{equation} \label{geneos} w_{\rm eff} = -1 +
\frac{2(n+2)}{3(2n+1)(n+1)} \ . 
\end{equation} 
Therefore, such modifications can
easily accommodate current observational
bounds~\cite{Melchiorri:2002ux,Spergel:2003cb} on the equation of
state parameter $-1.45< w_{\rm DE} <-0.74$ ($95\%$ confidence level).
In the asymptotic regime, $n=1$ is ruled out at this level, while
$n\geq 2$ is allowed; even $n=1$ is permitted if we are near
the top of the potential.

In this paper we seek to extend this approach. For the actions considered in this paper, a tranformation between the matter and Einstein frames does not necessarily make sense. Therefore, we would like to analyze the dynamics in the matter frame itself. A technique for this analysis is presented here. 

As an example, we begin by applying this technique to the CDTT model. We will work in the matter frame with a spatially flat Robertson-Walker metric
\begin{equation}
ds^2=-dt^2+a(t)^2\,d\vec x^2\ ,
\end{equation}
with $a(t)$ being the scale factor. The time-time component of the field equations, the {\it Friedmann equation} in this non-standard cosmology, is
\begin{equation}
\label{newfriedmann}
3H^2 - \frac{\mu^4}{12({\dot H}+2H^2)^3}\left(2H{\ddot H}
+15H^2{\dot H}+2{\dot H}^2+6H^4\right) = \frac{\rho_M}{\mpl^2}\ ,
\end{equation}
where an overdot denotes differentiation with respect to cosmic time,
and $H=\dot a/a$.

To perform a phase-space analysis of such equations we write ${\dot H}$ as a function of $H$. Let
\begin{eqnarray}
x &=& -H(t) \nonumber \\
y &=& {\dot H}(t) \ ,
\end{eqnarray}
so that
\begin{equation}
\ddot H =\frac{d\dot H}{dt} = \frac{d\dot H}{dH}\,\frac{dH}{dt} = - y\,\frac{dy}{dx}\ .
\end{equation}

In this way we may write~(\ref{newfriedmann}), a third order equation in $a(t)$, as a
second order equation in $H(t)$ and hence as a first order equation in $y(x)$.
The fact that the FRW equation for any $f(R)$ theory can be reduced
to the first-order equation for the spatially flat case (second order in case of a non-zero
spatial curvature) was first introduced in \cite{Gurovich:1979xg}, \cite{Starobinsky:1980te} .

Many cosmologically interesting solutions, including accelerating ones, are power-law solutions of 
the form $a(t)\propto t^p$. In such cases ${\dot H}=-H^2/p$ (i.e. $y=-x^2/p$). In anticipation of finding such solutions as asymptotic solutions to our equations, we define a new function $v(x)$ by
\begin{equation}
v(x) = -\frac{x^2}{y} \ ,
\end{equation}
with $v\neq 0$. Power-law solutions in the asymptotic future are then easily identified as  $v(x)\rightarrow p= {\rm constant}$ as $H\rightarrow 0$.

Furthermore, if
\begin{equation}
|\dot H|=|y|\to\infty ,\qquad{\rm then}\qquad |v|\to 0\quad
\textrm{if $x\neq0$.}
\end{equation}
So if $x$ is not zero, as $v\to0$ we approach the singularity $|\dot H|\to\infty$. This trick is invoked throughout this paper.

\begin{figure}[ht]
{\centering \includegraphics[width=3in]{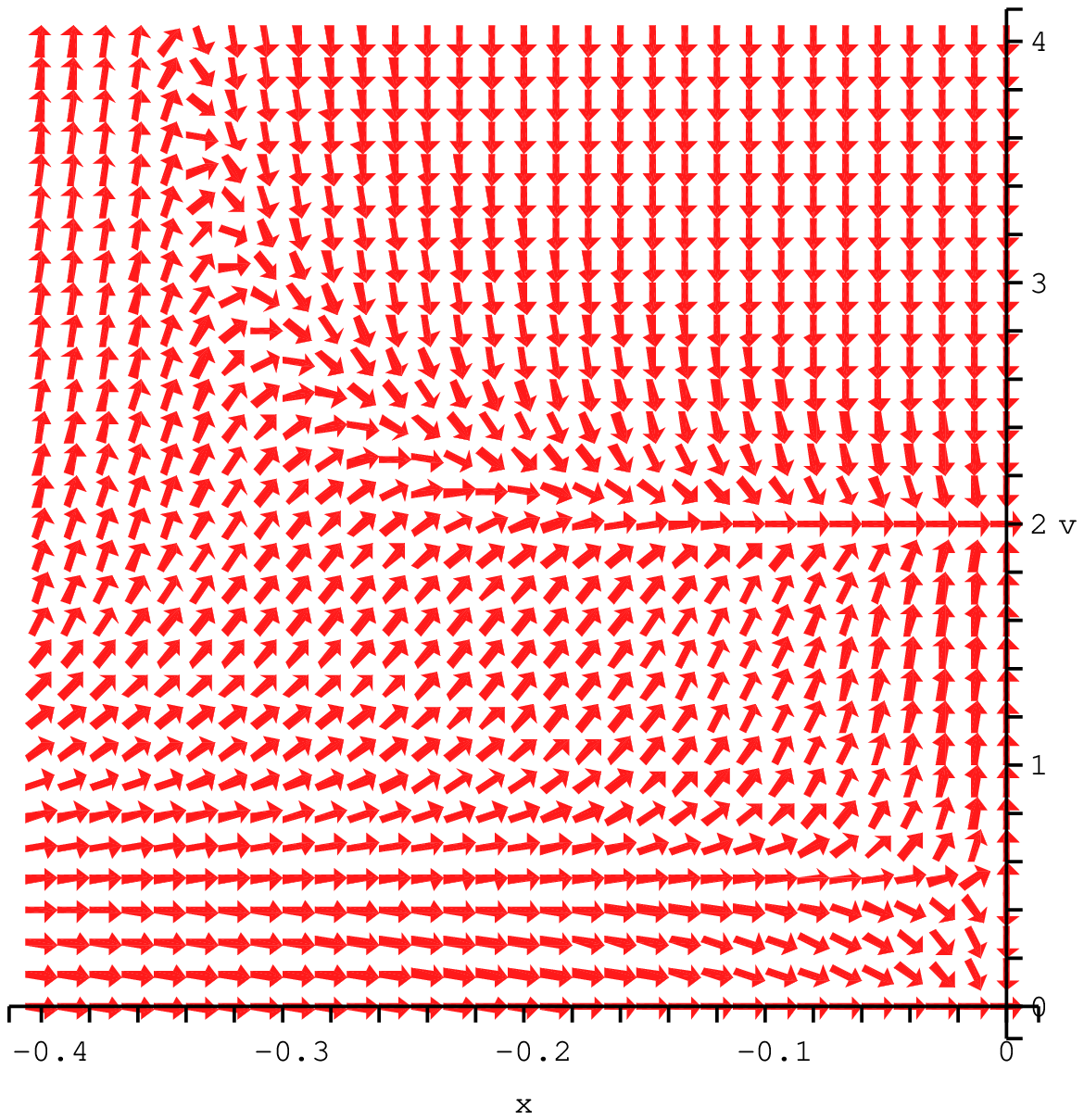} \includegraphics[width=3in]{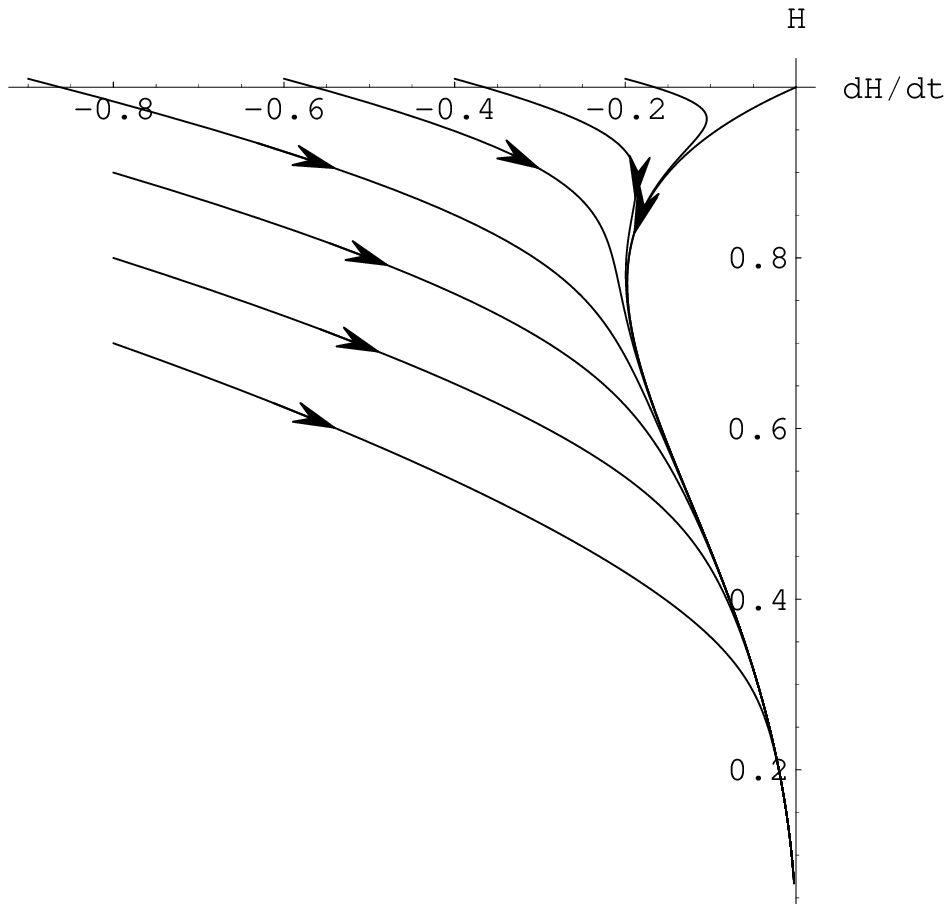} \par }
\caption{Two phase portraits for the modified gravity model proposed in~\cite{Carroll:2003wy}.
The left portrait is in the coordinates $(x,v)$, for which an attractor at constant $v=p$
corresponds to a power-law solution with $a(t)\propto t^p$. The right portrait is for the same 
theory in the $({\dot H},H)$ plane, with the unstable de Sitter solution at $(0,1)$.}
\label{Rmodelfig}
\end{figure}

Here, let us apply it to the simple case of~(\ref{newfriedmann}). The relevant first order equation is
\begin{equation}
x\,\frac{dv}{dx}=2v + \frac1{2\mu^4}
\left[x^4\,(36 - 216 v +432 v^2 -288 v^3) + \mu^4\,(2 v -15 v^2 + 6 v^3)\right],\label{eq_R}
\end{equation}
with the resulting phase plot shown in figure~\ref{Rmodelfig}.

Since the $x$-axis on this plot is (minus) the Hubble parameter, earlier times in the universe lie to the (negative) left and late times lie closer to $x=0$ (in all but exponential or phantom evolution).

Note that the numerical solution shows the accelerating attractor $a(t)\propto t^2$, corresponding to $v(x)\rightarrow 2$ as $x\rightarrow 0$, as expected from the analytic, Einstein-frame method. Indeed, 
for general $n$ this attractor, at $v=(2n+1)(n+1)/(n+2)$ can be obtained directly from the asymptotic form of the generalization of equation~(\ref{eq_R}). In addition, our method pinpoints a singularity in the phase space, corresponding to a power-law evolution with exponent $p=1/2$ (that of radiation). Both features are also evident from the study of the asymptotically late time behavior of equation~(\ref{eq_R}). In order to have a constant $v_0$ as a
solution, we need
\begin{equation}
2v_0^2-5v_0+2=0\ ,
\end{equation}
which has two real solutions: $v_0=1/2,2$, as expected.

The singularity at $p=1/2$ occurs because the Ricci scalar vanishes for this particular power. One might worry that this is problematic for describing the radiation-dominated phase in standard cosmology, since nucleosynthesis occurs during this epoch and provides a particularly strong constraint on deviations from the standard Friedmann equation at that time. However, as we shall see later, this is not a problem when matter sources are included explicitly, since even during radiation domination the Ricci scalar does not vanish exactly, but rather, has a small contribution from non-relativistic matter. 

Nevertheless, the singularity is a new feature that was not found in the Einstein-frame analysis~\cite{Carroll:2003wy}. This is presumably because  $R=0$ is a singular point of the conformal transformation used to reach the Einstein frame. 
We shall see similar singularities for some of the more general actions we consider in this paper.

Another way to visualize the solutions to our models is to use a more traditional phase portrait in the $({\dot H},H)$ plane. 

The outline of this paper is as follows. In the next section we shall introduce the general class of actions
we are interested in. In section~\ref{sec:vacuum} we analyze the vacuum equations, describing the singularity
and attractor structure in detail before moving on to some simple special cases. In 
section~\ref{sec:matter} we introduce matter into the equations, demonstrating briefly that the late-time
attractor solutions of the system remain unchanged and setting up the formalism used in the appendix
to establish stability of the system. In section~\ref{sec:conclusions} we summarize our findings and comment
on the status of these models as origins of cosmic acceleration. The paper contains two appendices. 
Appendix A contains definitions of a number of functions used in the body of the paper and Appendix B
consists of a proof of the stability of the vacuum solutions under the addition of matter.

\section{A General New Gravitational Action}
\label{sec:newaction}
We now generalize the action of~\cite{Carroll:2003wy} to include other curvature invariants.
There are, of course, any number of terms that we could consider. We have chosen to consider 
those invariants of lowest mass dimension that are also parity-conserving
\begin{eqnarray}
P &\equiv & R_{\mu\nu}\,R^{\mu\nu} \ \nonumber \\
Q &\equiv & R_{\alpha\beta\gamma\delta}\,R^{\alpha\beta\gamma\delta} \ .
\end{eqnarray}
Since we are interested in adding terms to the action that explicitly forbid flat space as a solution,
we will, in a similar way as in~\cite{Carroll:2003wy}, consider inverse powers of the above invariants.

It is likely that such terms introduce ghost degrees of freedom. We shall not address this problem here, since it is beyond the scope of this paper. Rather, if ghosts arise we shall require that some as yet unknown mechanism (for example, extra-dimensional effects) cut off the
theory in such a way that the associated instabilities do not appear on cosmological time scales~\cite{thanks} (see~\cite{Arkani-Hamed:2003uy} for an example of a concrete model where ghosts are brought under control by higher-derivative terms).
We therefore consider actions of the form
\begin{equation}
S=\int d^4x \sqrt{-g}\,[R+f(R,P,Q)] +\int d^4 x\, \sqrt{-g}\,
{\cal L}_M \ ,
\label{genaction}
\end{equation}
where $f(R,P,Q)$ is a general function describing deviations from general relativity.

It is convenient to define
\begin{equation}
f_R\equiv\frac{\partial f}{\partial R}\ , \qquad
f_P\equiv\frac{\partial f}{\partial P}\ , \qquad 
f_Q\equiv\frac{\partial f}{\partial Q} \ ,
\end{equation}
in terms of which the equations of motion are
\begin{eqnarray}
R_{\mu\nu} &-& \tfrac12\,g_{\mu\nu}\,R-\tfrac12\,g_{\mu\nu}\,f \nonumber \\
&+& f_R\,R_{\mu\nu}+2f_P\,R^\alpha{}_\mu\,R_{\alpha\nu}
+2f_Q\,R_{\alpha\beta\gamma\mu}\,R^{\alpha\beta\gamma}{}_\nu\nonumber\\
&+& g_{\mu\nu}\,\Box f_R -\nabla_\mu\nabla_\nu f_R
-2\nabla_\alpha\nabla_\beta[f_P\,R^\alpha{}_{(\mu}{}\delta^\beta{}_{\nu)}]
+\Box(f_P\,R_{\mu\nu})\nonumber\\
&+& g_{\mu\nu}\,\nabla_\alpha\nabla_\beta(f_P\,R^{\alpha\beta})
-4\nabla_\alpha\nabla_\beta[f_Q\,R^\alpha{}_{(\mu\nu)}{}^\beta]
=8\pi G\,T_{\mu\nu}\ .\label{equaz}
\end{eqnarray}

It is straightforward to check that these equations reduce to those of the simple model of~\cite{Carroll:2003wy} 
for $f(R)=-\mu^4/R$.

We would like to obtain constant curvature vacuum solutions to these field equations. To do so, we take the trace of~(\ref{equaz}) and substitute $Q=R^2/4$ and $P=R^2/6$ (which are identities satisfied by constant curvature spacetimes) into the resulting equation to obtain the algebraic equation:
\begin{equation}
\label{constcurvexp}
\left(2f_{Q}+3f_{P}\right)R^2+6\left(f_R -1\right)R -12f=0 \ .
\end{equation}
Solving this equation for the Ricci scalar yields the constant curvature vacuum solutions.

Evidently, actions of the form~(\ref{genaction}) generically admit a maximally-symmetric solution: 
$R=$ a non-zero constant. However, an equally generic feature of such models is that this deSitter solution is unstable. In the CDTT model the instability is to an accelerating power-law attractor. This is a possibility that we will also see in many of the more general models under consideration here.

Before we leave this section, note that, in analyzing the cosmology of these general models, it is useful to have at hand the following expressions, which hold in a flat FRW background
\begin{eqnarray}
R&=& 6\left(\frac{\dot a^2}{a^2}+\frac{\ddot a}a\right) =6\,(\dot H + 2 H^2) \\
P&=&12\left(\frac{\dot a^4}{a^4}+\frac{\ddot a^2}{a^2}+
     \frac{\dot a^2}{a^2}\,\frac{\ddot a}a\right) =12\,[(\dot H + H^2)^2 + H^4+H^2\,(\dot H + H^2)]  \\
Q&=&12\left(\frac{\dot a^4}{a^4}+\frac{\ddot a^2}{a^2}\right) =12\,[(\dot H + H^2)^2 + H^4] \ .
\end{eqnarray}
We have provided these both in terms of the scale factor $a(t)$ and in terms of the Hubble parameter $H(t)={\dot a}(t)/a(t)$, since they will be separately important in this paper.

\section{Vacuum Solutions}
\label{sec:vacuum}
In this section we study cosmological solutions to the field equations~(\ref{equaz}) in the absence of matter sources. 
Physically, this is important because we are hoping to find novel cosmological consequences arising purely from
the gravitational sector of the theory. Mathematically, this provides us with valuable insight into the structure of the equations, which take a significantly simplified form wherein the Hubble parameter may be treated as the independent variable.

As mentioned in the previous section, we will consider inverse powers of our curvature invariants and, for 
simplicity, we will specialize to a class of actions with
\begin{equation}
f(R,P,Q)=-\frac{\mu^{4n+2}}{(aR^2+b P+c Q)^n} \ ,
\label{f_R}
\end{equation}
where $n$ is a positive integer, $\mu$ has dimensions of mass and $a$, $b$ and $c$ are dimensionless constants.
$^{\footnotemark[1]}$
\footnotetext[1]{Another potentially interesting possibility is a correction of form $f(R,P,Q)=\frac{\mu ^4 R}{P}$. This term has the same mass dimension as the $\mu ^4/R$ term of CDTT. However, it turns out that this model does not possess any accelerating attractors. Specifically, the scale factor asymptotically approaches $a(t) \propto t^{0.37}$.}
We will focus on the case $n=1$ for most of the paper, because the analysis is less involved for that case. 
For general $n$ the qualitative features of the system are as for $n=1$ and we discuss the quantitative
differences in our conclusions.

\subsection{Distinguished Points of the Action (and Equations)}
In general, the analogue of the Friedmann equation may be written in
the following convenient form
\begin{equation}
\frac{A+B\ddot H}C=M \ ,
\label{compact_eom}
\end{equation}
where $A=A(H,\dot H)$, $B = B(H,\dot H)$ and $C=C(H,\dot H)$ arise
from the gravitational part of the action and $M=M(a)$ describes the
possible inclusion of matter. 

It is also convenient to write this schematically in terms of our
variables of the previous section as
\begin{equation}
x\,\frac{dv}{dx}=\frac{x^6\,f(v)+\mu^6\,v^2\,g(v)}{2\mu^6\,v\,h(v)}\ ,
\label{eqa}
\end{equation}
where $f(v)$, $g(v)$ and $h(v)$ are 6th, 4th and 2nd degree
polynomials respectively in the variable $v$, whose explicit form is given in Appendix A. We are often interested
in a particular subset of the phase space, the region where $v>1/2$. This is because from nucleosynthesis to the matter-dominated epoch, we expect Einstein's equations to provide a good approximation to the dynamics, and therefore, when matter becomes subdominant we should have $1/2<v<2/3$. 

There are three types of special points in the phase space plots of
these equations:

\begin{enumerate}
\item {\it Singular Points of the Friedmann Equation}  

In our introduction we
reconsidered the model of~\cite{Carroll:2003wy} and discovered a
singular point of both the action and the equations of motion,
corresponding to a power-law evolution with exponent $p=1/2$. In this
particular case the singularity occurred because $R\equiv0$ for
$p=1/2$. 
In our more general models, analogous singularities occur
whenever the denominator of the Friedmann equation blows up; i.e. at
zeros of $C$, where $C$ is defined by (\ref{compact_eom}). For the flat cosmological ansatz this occurs when
\begin{equation}
{\dot H}+H^2=-\frac{\alpha}{4}\left(\frac{{\dot H}}{H}\right)^2 \ ,
\end{equation}
where we have defined
\begin{equation}
\alpha\equiv \frac{12a+4b+4c}{12a+3b+2c} .
\end{equation}
Recall that $a,b,$ and $c$ are parameters in the Lagrangian (\ref{f_R}). In our variables of the previous section this becomes
\begin{equation}
v=p_{1,2} \equiv \frac{1}{2}\left(1\pm\sqrt{1-\alpha}\right) \ ,
\end{equation}
each zero having multiplicity 3.

These singularities only exist if $p_{1,2}$ are real, i.e.\ $\alpha\leq 1$ and
therefore there are many invariants that do not admit this type of
singularity. In the simple case of~\cite{Carroll:2003wy}, corresponding to
$b=c=0$, note that we have $\alpha=1$ and recover $p_c=1/2$ as expected. If
$\alpha\leq1$
\begin{equation}
p_1\leq1/2\qquad p_2\geq1/2
\end{equation}

\item  {\it Singular Points at which} $\ddot H \rightarrow \infty$
 
Points at which $\ddot H \rightarrow \infty$ occur at zeros of $B$ and, in our  variables of the 
previous section, correspond to
\begin{equation}
\left|\frac{dv}{dx}\right| \rightarrow +\infty
\end{equation}
at finite $x$ and $v$. We denote the zeros by $v(x)=v_1,v_2$, where the $v_i$ are (in general
complex) constants constructed from $a$, $b$ and $c$.

If $g(v)\neq0$ one singular point is at $x=0$. Otherwise the singularities occur at
solutions of $v\,h(v)=0$, i.e.
\begin{eqnarray}
v\left[(108a^2\right.&+&51ab+7bc+30ac +2c^2+6b^2)\,v^2 \nonumber \\
&-&(63ab+6c^2+9b^2+108a^2+15bc+54ac)\,v\nonumber \\
&&\left.\ \ \ \ \ \ \ \ \ \ +18ab+3c^2+6bc+27a^2+18ac+3b^2\right]=0\ ,
\end{eqnarray}
which are given by
\begin{eqnarray}
v_{1,2} &=& \frac{3\alpha}{2(4-\alpha)}\left(1\pm \sqrt{\frac{\alpha -1}{3}}\right) \nonumber \\
v_3&=&0 \ .
\end{eqnarray}

\item {\it Late time Stable Points.} 

Finally, we look for late time power-law attractors. This happens if $v(x)\rightarrow$ constant (distinct
from our previously mentioned singular values). Taking an asymptotic limit of the equations of motion yields
\begin{eqnarray}
v\left[(288a^2\right. &+& 18b^2+8c^2+144ab+96ac+24bc)\,v^4 \nonumber \\
&-&(1512a^2+90b^2+36c^2+738ab+468ac+114bc)\,v^3 \nonumber \\
&&\ \ \ \ \ +(1836a^2+123b^2+62c^2+951ab+678ac+175bc)\,v^2 \nonumber \\
&&\ \ \ \ \ \ \ \ \ \ -(846a^2+69b^2+44c^2+489ab+414ac+113bc)\,v \nonumber \\
&&\left. \ \ \ \ \ \ \ \ \ \ \ \ \ \ \ +135a^2+15b^2+15c^2+90ab+90ac+30bc\right]=0 \ ,
\end{eqnarray}
with solutions
\begin{eqnarray}
s_{1,2} &=&  \frac{20-3\alpha}{8}\left(1\pm \sqrt{1-\frac{120\alpha}{(20-3\alpha)^2}}\right) \nonumber\\
s_3&=&p_1\\
s_4&=&p_2\\
s_5&=&0\ ,
\end{eqnarray}
and $s_2>s_1$.

It is clear that only $s_1$ and $s_2$ can be late-time power-law
solutions. The other ones represent singularities; $v=0$ implies that
$y=\dot H\rightarrow \infty$, whereas $s_3$ and $s_4$ are the singular points we discussed earlier. 

\end{enumerate}

\subsection{Summary of Possibilities}
Here we summarize the different vacuum possibilities. It is useful to define the following two constants
\begin{eqnarray}
\eta_1=\frac{280+60\sqrt{3}}{111+60\sqrt{3}}\approx 1.78 \\
\eta_2=\frac{280-60\sqrt{3}}{111-60\sqrt{3}}\approx 24.9 \ .
\end{eqnarray}

\begin{enumerate}
\item \underline{$\alpha<1$}. In this case $v_i$ are complex, whereas $p_i$
and $s_i$ are real. It is straightforward to show that $s_2>p_2>1/2$ and
$s_2>1$.
\begin{itemize}
\item $0<\alpha<1$. In this case $1/2>s_1>p_1>0$ and $1>p_2>1/2$. 
Solutions close to $p_2$ are repelled
from it, whereas $s_1$ is an attractor. This leads to decelerating late time behavior.
\item $0>\alpha>-160/9$. Here $s_1<p_1<0$ and solutions are attracted to
  $(x=0,v=0)$. Furthermore $p_2>1$.
\item $\alpha<-160/9$. In this case $p_1<s_1<0$ and again we have that
solutions are attracted to $(x=0,v=0)$. Again, $p_2>1$.
\end{itemize}

\item \underline{$\alpha>1$}. In this case $v_i$ are real, whereas $p_i$ are complex.
\begin{itemize}
\item $1<\alpha<4/3$. Here $v_1<1/2$ and $1/2<s_1<v_2<1$. Furthermore
$s_2>1$. Both $s_1$ and $s_2$ are attractors, but $s_1$ describes a
decelerating phase.
\item $4/3<\alpha<\eta_1$. Here $1/2<v_1<1$ and $v_2>1$. Both $(x=0,v=0)$ and
$s_{1,2}$ are attractors, with the separatrix being at $v_{1,2}$. 
\item $\eta_1 <\alpha<4$. In this case $1/2<v_1<1$ and
$v_2>1$. Again, both $(x=0,v=0)$ and $v_2$ are attractors and the separatrix is
at $v_1$. There are no real solutions for $s_{1,2}$.
\item $4<\alpha<\eta_2$. Here there are no real late-time attractors since the $s_i$ are
complex. We also have $v_2<0$ and $v_1>1$. Solutions either evolve to $(x=0,v=0)$ or 
to $v=+\infty$, with separatrix at $v_1$.
\item $\alpha>\eta_2$. In this case $v_2<0$, $v_1>1$ and
$s_1<s_2<0$. Evolution is to a decelerating attractor.
\end{itemize}

\item \underline{$\alpha =1$}. This yields a promising class of solutions. 
We have
\begin{equation}
p_1=p_2=v_1=v_2=s_1=\frac12\ .
\end{equation}
All singularities occur as $v(x) \rightarrow 1/2$. However, from nucleosynthesis onwards we never encounter 
this point. It is simple to show that solutions evolve to a late-time power-law
attractor describing an accelerating phase, with
\begin{equation}
s_2=\frac{15}4=3.75\ ,
\end{equation}
which is otherwise independent of $a,b,c$.
\end{enumerate}

\begin{figure}[ht]
{\centering \includegraphics[width=4in]{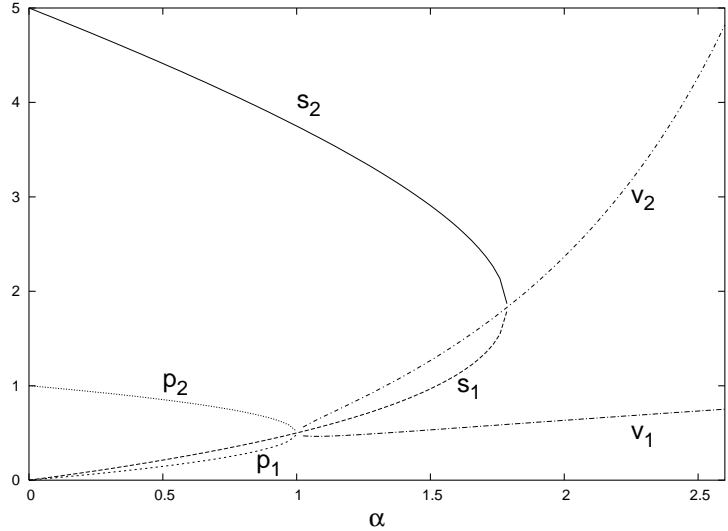} \par}
\caption{The values of the various distinguished points as $\alpha$ is varied.}
\label{studio-sing}
\end{figure}

The results of this section may be summarized in figure~\ref{studio-sing}, 
showing the values of the various distinguished points as $\alpha$ is varied.

\subsection{Inverse Powers of $P\equiv  R_{\mu\nu}\,R^{\mu\nu}$}
Let us begin by dealing only with actions containing modifications
involving $P\equiv R_{\mu\nu}\,R^{\mu\nu}$. Our prototype is
to consider $f(P)=-m^6/P$, with $m$ a parameter with dimensions
of mass.

Using~(\ref{constcurvexp}) we can see that there is a constant curvature vacuum solution to this action given by
\begin{equation}
R_{\rm const}^{(P)} =\left(16\right)^{1/3}m^2 \ .
\end{equation}
However, we would like to investigate other cosmological solutions and analyze their stability.
 
From~(\ref{equaz}), with the flat cosmological ansatz, the analogue of the Friedmann equation becomes
\bea
3H^2&-&\frac{m^6}{8(3H^4+3H^2\dot H + \dot H^2)^3} 
\left[\dot H^4 + 11 H^2 \dot H^3 + 2 H \dot H^2\ddot H 
\right. \notag \\
&&\left.{}+33 H^4\dot H^2+
 30 H^6 \dot H +6 H^3 \dot H \ddot H +
 6 H^8 +  4 H^5\ddot H\right] =0 \ .
\label{riccieqn}
\eea

\begin{figure}[ht]
{\centering \includegraphics[width=3in]{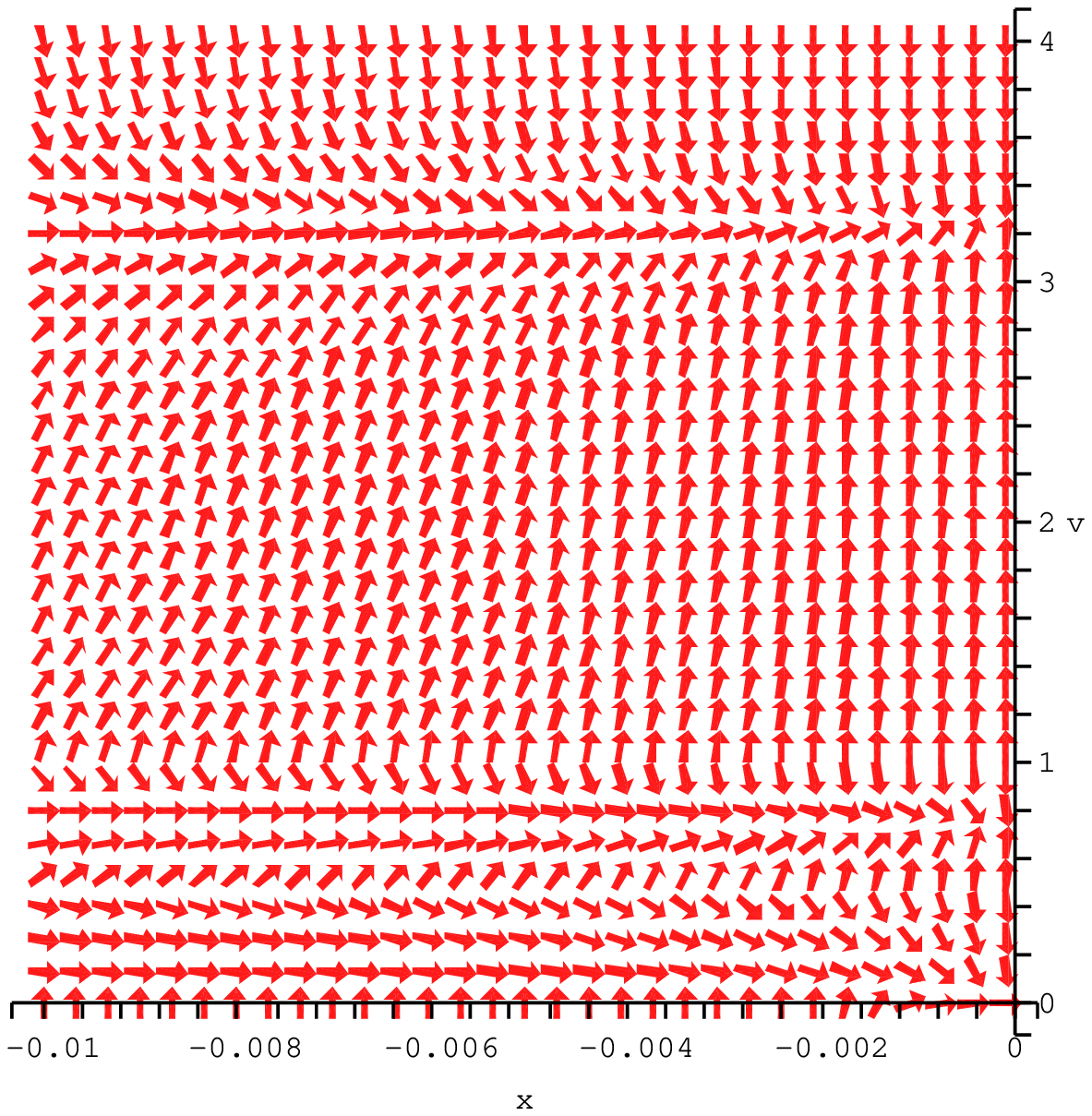} \includegraphics[width=3in]{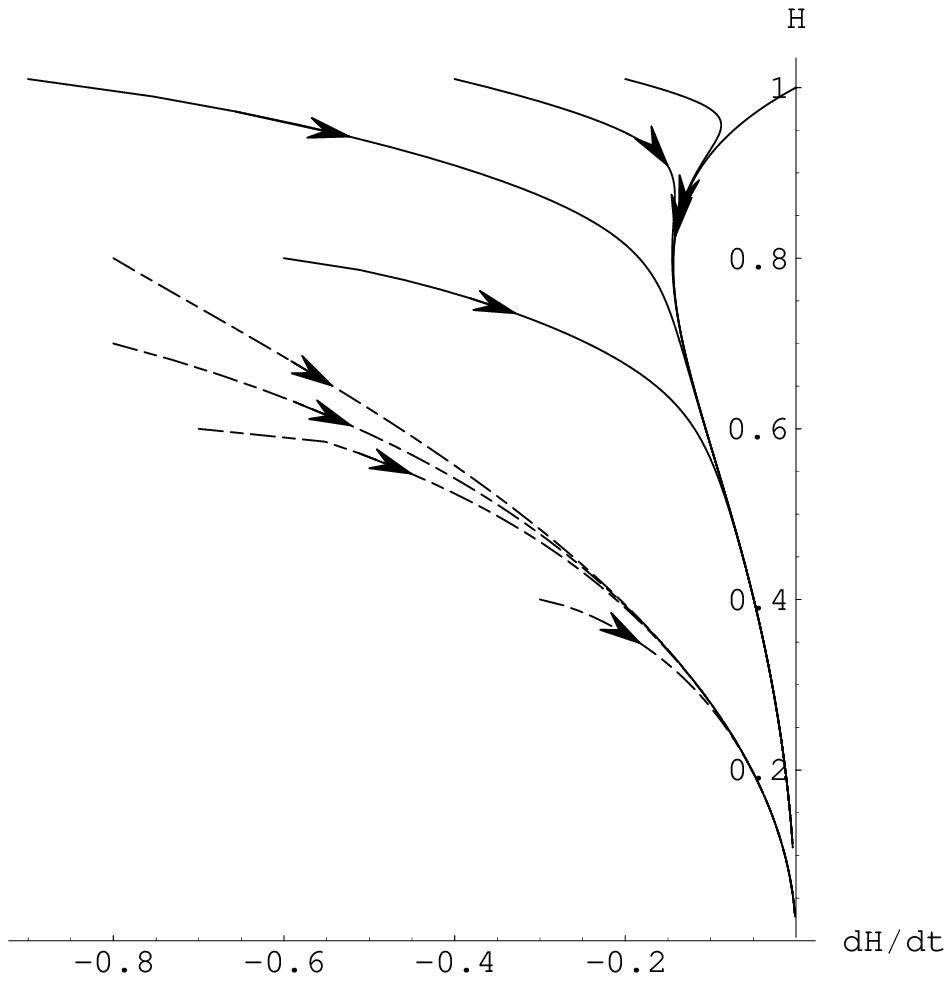} \par}
\caption{Two phase portraits for the $f(R,P,Q)=-m^6/P$ modification.
The left portrait is in the coordinates $(x,v)$, for which an attractor at constant $v=p$
corresponds to a power-law solution with $a(t)\propto t^p$. The right portrait is for the same 
theory in the $({\dot H},H)$ plane. There are two 
late-time power law attractors corresponding to $p=(4\pm \sqrt{6})/2$.
The (-) branch, represented by the solid lines, is non-accelerating, while the (+) branch,
represented by the dash-dotted line, is accelerating.}
\label{ricciphaseplot}
\end{figure}

We analyze this equation using the same technique, with the same definitions, as in the example of the previous section. The relevant equation is 
\begin{eqnarray}
x\,\frac{dv}{dx}=2v^2 &+& \frac{1}{2m^6 v (2v^2-3v +1)}\left[-x^6(24 - 216 v + 864 v^2 - 1944v^3\right. \nonumber\\
&+& \left. 2592 v^4 - 1944 v^5 + 648 v^6)+m^6 \,(v^2 - 11 v^3 + 33 v^4 - 30 v^5 + 6 v^6)\right]\ .
\label{eq_ricci}
\end{eqnarray}
The solution to this equation is displayed graphically in figure~\ref{ricciphaseplot}. 
We identify four fixed points of the system; two attractors at $v\simeq 0.77$ and $v\simeq 3.22$ and two
repellers at $v\simeq 0.5$ and $v=1$. Clearly, in order to obtain a late-time accelerating solution ($p>1$), it is necessary to give accelerating initial conditions (${\ddot a}>0$), otherwise the system is in the basin of attraction of the non-accelerating attractor at 
$p\simeq 0.77$.

The exact exponents of the two late-time attractors of the system are obtained
by studying the asymptotic behavior of~(\ref{eq_ricci}). Substituting in a
power-law ansatz and taking the late-time limit we find that, in order to have
a constant $v=v_0$ as a solution, the exponent must satisfy
\begin{equation}
6v_0^4-30v_0^3+41v_0^2-23v_0+5=0\ .
\end{equation}
This equation has two real solutions (and two complex ones). The real
solutions are: $v_0=2-\sqrt{6}/2\simeq 0.77$ and $v_0=2+\sqrt{6}/2\simeq 3.22$.

Even the non-accelerating attractor is of some interest in this model. In
order for structure to form in the universe, there must be a sufficiently long
epoch of matter domination, for which $a(t)\propto t^{2/3}$. As matter
redshifts away, however, since the universe is decelerating, we expect the
universe to approach the attractor at $p\simeq 0.77$. This corresponds to an
effective equation of state $w_{\rm eff}\simeq -0.13$; i.e. negative pressure,
although not negative enough to provide a good fit to the supernova
observations.

\subsection{Inverse Powers of $Q \equiv R_{\alpha\beta\gamma\delta}\,R^{\alpha\beta\gamma\delta}$}

Now let us move on to actions containing modifications involving only
$Q = R_{\alpha\beta\gamma\delta}\,R^{\alpha\beta\gamma\delta}$. Our prototype
example is $f(Q)=-M^6/Q$, with $M$ another parameter with dimensions of mass, 
and the analysis follows much the same as in the previous subsection,
albeit with different results.

Again,~(\ref{constcurvexp}) demonstrates that there is a constant curvature vacuum solution to this action given by
\begin{equation}
R_{\rm const}^{(Q)} =\left(24\right)^{1/3}M^2 \ .
\end{equation}

What about other possible cosmological solutions?
From~(\ref{equaz}), with the flat cosmological ansatz, the analogue of the Friedmann equation becomes
\bea
3H^2&-&\frac{M^6}{24[\dot H^2 + 2 H^2 \dot H + 2 H^4]^3} \left[8 H^8 + 36  H^6 \dot H +54  H^4 \dot H^2 \right.\notag\\
&&\left.{}+ 20 H^2 \dot H^3+ 3 \dot H^4
+ 4  H^5 \ddot H + 12  H^3 \dot H \ddot H + 
6  H \dot H^2 \ddot H\right] = 0\ .
\eea
Employing our phase space technique once more, in the variables best-suited for analyzing power-law behavior, the relevant
equation is
\begin{align}
x\,\frac{dv}{dx}&=2v + \frac{1}{2M^6 v [2v^2 - 6 v + 3]}\left[-x^6 (576 v^6 - 1728 v^5 + 2592 v^4 - 2304 v^3
\right.  \notag\\
&+ \left.1296 v^2 - 432 v +72)+M^6\,(8v^6 -36 v^5 + 54 v^4 - 20 v^3 + 3v^2)\right] \ .
\label{eq_rieman}
\end{align}
It is clear from this that our action consisting of $f(R,P,Q)=-M^6/Q$ does not admit any late-time power-law
attractors.
\begin{figure}[ht]
{\centering \includegraphics[width=3.5in]{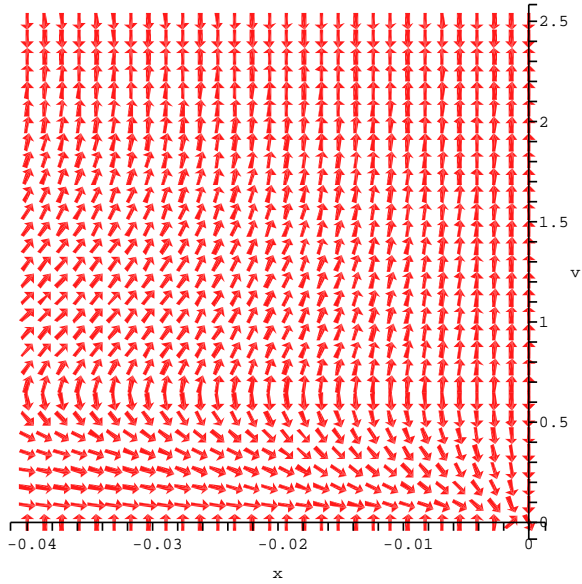} \par}
\caption{Phase portrait for the  $f(R,P,Q)=-M^6/Q$ modification in the coordinates 
$(x,v)$, for which an attractor at constant $v=p$
corresponds to a power-law solution with $a(t)\propto t^p$.}
\label{Riemannfigure}
\end{figure}

This is consistent with a study of the late-time asymptotics of~(\ref{eq_rieman}). As in our previous
analysis, late-time power-law solutions correspond to real solutions to the equation
\begin{equation}
8v_0^4-36v_0^3+62v_0^2-44v_0+15=0\ .
\end{equation}
However, no such real solutions exist, confirming our phase-space analysis.

\section{Including Matter}
\label{sec:matter}
We now show that the late-time behavior of our vacuum solutions remains unaltered upon the inclusion of matter. 

We begin by rewriting the equation of motion in a more convenient form. Let $\Sigma_{\mu \nu}$ be the tensor defined by the left hand side of~(\ref{equaz}). Then, the generalized Friedmann equation takes the form
\begin{equation}
\Sigma_{00}=8\pi G \rho ,
\label{mateq1}
\end{equation}
where $\rho$ is the energy density of a perfect fluid with equation of state $p=w\rho$. Now, since $x=-H$, $y=-\dot{x}$, and $\rho \propto a^{-3(1+w)}$, we have $\frac{d\ln (\rho/\rho_0)}{dx}=3(1+w)\frac{v}{x}$. Combining this relation with~(\ref{mateq1}) yields the equation of motion
\begin{equation}
\label{mateq2}
x\frac{d\Sigma_{00}}{dx}=3(1+w)v\Sigma_{00}.
\end{equation}

Thus far, we have not imposed any dynamics. Specializing to a theory with $f(R,P,Q)=\mu ^6 / (a R^2 +b P + c Q)$ gives
\begin{equation}
\label{mateq3}
\Sigma_{00} = \frac1{24\,x^4}\,F(x,v,s) ,
\end{equation}
with
\begin{equation}
\label{defF}
F = 72\,x^6+\mu^6\,\frac{v^2\,g(v)-2\,x\,v\,h(v)\,s}{[d(v)]^3} ,
\end{equation}
where $s=\frac{dv}{dx}$. The explicit forms of the functions $h(v), g(v)$, and $d(v)$, defined in Appendix A, are not needed in this section.  Substituting~(\ref{mateq3}) and~(\ref{defF}) into~(\ref{mateq2}) gives the equation governing the dynamics of our theory in the presence of matter:
\begin{equation}
x\,\frac{dF}{dx}=\gamma\,(v)\,F\ ,
\label{seco}
\end{equation}
where $\gamma (v)\equiv [3(1+w)\,v+4]\,$.
Using the chain rule this becomes
\begin{equation}
x(F_x+F_v\,s+F_s\,s')=\gamma(v)\,F\ ,
\label{linea}
\end{equation}
where
\begin{eqnarray}
F_x&=&\frac{\partial F}{\partial x}=
432\,x^5-2\mu^6\,\frac{v\,h\,s}{d^3}\\
F_v&=&\frac{\partial F}{\partial v}=
\mu^6\,\frac{v^2\,g_v+2\,v\,g-2\,x\,h\,s-
2\,x\,v\,h_v\,s}{d^3}-3\,\mu^6\,\frac{v^2g-2xvhs}{d^4}\,d_v\nonumber\\
F_s&=&\frac{\partial F}{\partial s}=-\frac{2\mu^6\,x\,v\,h}{d^3}
\end{eqnarray}
and a prime denotes differentiation with respect to $x$.

Seeking late-time power law behavior, we take the limits $x\rightarrow 0$ and $s\rightarrow 0$ in~(\ref{linea}). This yields the condition $g(v_0)=0$, which is solved by the same power law fixed points as those obtained in vacuum. This is as one might expect, since matter redshifts away in the asymptotic
future. However, the above description proves useful for dealing with the issue of stability. Since this 
is somewhat technical, we relegate the details to Appendix B and merely assert here that these fixed points 
do remain stable in the presence of matter sources.

\section{Comments and Conclusions}
\label{sec:conclusions}
In the extreme low curvature regime, our only tests of general relativity are cosmological. The discovery
of new phenomena at these scales may point to new matter sources, but alternatively may
hint at hitherto undetected modifications of gravity. 

The acceleration of the universe provides a particular challenge to modifications of gravity. Unlike the
known perturbative corrections to the Einstein-Hilbert action arising from string theory, late time 
acceleration requires modifications that become important at extremely low energies, so low that only
today, at the largest scales in the universe, is the resulting curvature low enough to lead to measurable 
deviations from general relativity. 

These considerations led some of the current authors and others to consider new terms in the action 
for gravity that consist of inverse powers of the Ricci scalar. It is easy to show that such an approach
introduces a de Sitter solution. However, this solution is unstable to a late-time accelerating power
law attractor. For appropriate choices of parameters, this theory is a candidate to explain cosmic
acceleration without the need for dark energy, although the simplest such theories are in conflict
with solar system tests. For Lagrangians that are functions of only the Ricci scalar, there exists a map
to an Einstein frame, in which the new degrees of freedom are represented by a scalar field. As a result,
such modified gravity theories share many features in common with some dark energy models.

In this paper we have introduced a much more general class of modifications to the Einstein-Hilbert action,
becoming relevant at extremely low curvatures. Specifically, we have considered inverse powers of 
arbitrary linear combinations of the curvature invariants $R^2$, $P\equiv  R_{\mu\nu}\,R^{\mu\nu}$ and
$Q \equiv R_{\alpha\beta\gamma\delta}\,R^{\alpha\beta\gamma\delta}$. Such modifications are not 
simply equivalent to Einstein gravity plus scalar matter sources.

We have performed a general analysis of the late-time evolution of cosmological solutions to these 
theories.  Many of the theories exhibit late-time attractors of the form $a(t)\propto t^p$, with $p$ some
constant power. Indeed, there are often multiple such attractors. For a large class of theories there
exists at least one attractor satisfying $p>1$, corresponding to cosmic acceleration. 

The detailed structure of cosmological solutions to these theories turns out to be quite rich and varied,
depending on the dimensionless parameters entering the particular linear combination. Two distinct 
types of singularities may exist, as well as the late-time power law attractors. We have identified all
those theories for which the late time behavior is consistent with the observed acceleration of the
universe, providing a whole new class of theories - generalized modified gravity theories - which are alternatives to dark energy.

The results we have found for the modification $f=\mu^6/(a\,R^2+b\,P+c\,Q)$
may be generalized to modifications
\begin{equation}
f=\frac{\mu^{4n+2}}{(a\,R^2+b\,P+c\,Q)^n} \ ,
\end{equation}
using exactly the same calculational techniques.

In general there are power-law attractors with the following exponents, whenever they are real
\begin{equation}
v^{\rm gen}_{1,2}=\frac{8n^2+10n+2-3\alpha\pm
\sqrt\Gamma}{4(n+1)}\ ,
\end{equation}
where
\begin{eqnarray}
\Gamma&=&9n^2\alpha^2-(80n^3+116n^2+40n+4)\,\alpha\nonumber\\
&&\qquad{}+64n^4+160n^3+132n^2+40n+4 \ .
\end{eqnarray}
Clearly, as $n\rightarrow \infty$, the smaller attractor tends to 0, 
whereas the larger one increases linearly as $4n$.

Two special cases are
\begin{eqnarray}
f &=& -\frac{m^{4n+2}}{P^n} \ , \nonumber \\
f &=& -\frac{M^{4n+2}}{Q^n} \ ,
\end{eqnarray}
for which the power law attractors, $v_{1,2}$ are
\begin{eqnarray}
v^{(P)}_{1,2}&=&\frac{12n^2+9n+3\pm
\sqrt{144n^2+120n^3-15n^2-30n-3}}{2(3+3n)}
\label{VPN}\\
v^{(Q)}_{1,2}&=&\frac{4n^2+2n+1\pm
\sqrt{16n^4-16n^2-10n-1}}{2n+2}\ .
\label{VQN}
\end{eqnarray}

For $n=1$, all the above expressions agree with the values found
earlier. It is interesting to note that $v^{(Q)}_{1,2}$ are imaginary for $n=1$, but are real
for all $n>1$.

Of course, much remains to be done. We have not addressed solar system tests
of these theories since this is a complicated analysis that is beyond the
scope of our current paper.  
We have not focused on detailed comparisons between our models and the supernova data and it is possible that there are 
specific signatures of this new physics in such data. For example, the modified Friedmann equations arising from the theories presented here should directly provide 
information about the jerk parameter. We intend to consider such effects, not only in the present models, but in those 
proposed by us and other authors, in a separate paper focused on the connections of these models with observations. 
In this context, as with the case of the simple
modifications introduced in~\cite{Carroll:2003wy}, it may also be interesting
to study more complicated functions of the curvature invariants we have
considered.

\acknowledgments  
MT thanks Gia Dvali, Renata Kallosh, Burt Ovrut and Richard Woodard for helpful discussions.
The work of SMC is supported in
part by the Department of Energy (DOE), the NSF, and the Packard Foundation.  VD is supported
in part by the NSF and the DOE.  ADF, DAE and MT are supported in part by the NSF under grant
PHY-0094122. DAE is also supported in part by funds from Syracuse University and
MT is supported in part by a Cottrell Scholar Award from Research Corporation.  MST
is supported in part by the DOE (at Chicago), the NASA (at
Fermilab), and the NSF (at Chicago).

\appendix
\section{Some Definitions}
\label{app1}

In Section~\ref{sec:vacuum} we defined 

\begin{equation}
\alpha=\frac{4\,(3a+b+c)}\Delta\ ,
\end{equation}
where
\begin{equation}
\Delta=12a+3b+2c\ .
\end{equation}

We should distinguish betweeen two cases:

\subsection{$\Delta=0$}

In this case $\alpha$ diverges, but we may define 
\begin{eqnarray}
g(v)&=&(3a+b+c)\,(282a+69b+44c)\,v+15\,(3a+b+c)^2\\
h(v)&=&3\,(3a+b+c)^2\\
d(v)&=&3a+b+c\ .
\end{eqnarray}
Note that both $\Delta$, and $3a+b+c$, cannnot vanish simultaneously, as 
this causes the Friedmann equation to be singular for all values of $x$ and 
$v$.

\subsection{$\Delta\neq0$}
In this case we have 
\begin{eqnarray}
g(v) &=& 2\,\Delta^2\,(v-s_2)\,(v-s_1)\,(v-p_2)\,(v-p_1)\\
h(v) &=& \Delta^2\left(1-\frac\alpha4\right)(v-v_2)\,(v-v_1)\\
d(v) &=& \Delta\,(v-p_2)\,(v-p_1)\ ,
\end{eqnarray}
where either $v_{1,2}$ or $p_{1,2}$ are complex unless $\alpha=1$, in 
which case, $v_{1,2}=p_{1,2}=s_1=1/2$ and $s_2=3.75$. Also $s_{1,2}$ in general may 
be complex. As we saw in Section~\ref{sec:vacuum}, $s_i$, $v_i$ and $p_i$ are all
functions of $\alpha$ only.

\section{Stability of Fixed Points in the Presence of Matter}
\label{app2}
Let us rewrite (\ref{linea}) as a system of first-order ordinary differential equations
\begin{eqnarray}
x'&=&1\\ 
s'&=&\frac{\gamma(v)\,F}{x\,F_s}-\frac{F_v}{F_s}\,s-\frac{F_x}{F_s}\\
v'&=&s\ .
\end{eqnarray}
Treating $x$ as a dependent variable makes the system autonomous, facilitating a phase-space analysis.

\begin{figure}
\centering
\includegraphics[height=10truecm]{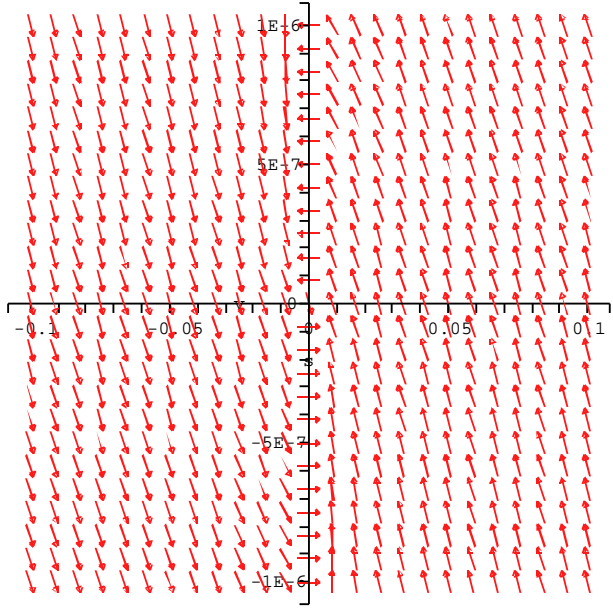}
\caption{Phase plot for the linearized equations close to the power
law solution in the coordinates 
$(x,v)$, for which an attractor at constant $v=p$
corresponds to a power-law solution with $a(t)\propto t^p$.}
\label{fig:linearized}
\end{figure}

To obtain the phase portrait of the system we define a vector field by
${\vec W}^T \equiv (x',s',v')$
and plot this in the vicinity of the fixed point. In figure~\ref{fig:linearized} we show a 2D slice of this phase portrait by choosing a section at constant $x$. Note that at the fixed point, $\vec W^{(0)}=(1,0,0)$.

To analyze the stability of the system, we first 
linearize the system about the fixed point $v=v_0$, $s=0$, and $x=x_0\ll1$ (all expressions evaluated at these values of $v,s$ and $x$ will carry the superscript $^{(0)}$) 

We write the linearized system of equations as
\begin{equation}
\label{matrixeq}
W_i=\sum_{\alpha}\left[M\right]_{i\alpha}^{(0)}(\eta_{\alpha}-\eta^{(0)}_{\alpha})+W_i^{(0)} \ ,
\end{equation}
the matrix $M$ being defined by
\begin{equation}
\left[M\right]_{i\alpha}^{(0)} =\left[\frac{\partial W_i}{\partial \eta _{\alpha}}\right]^{(0)} \ ,
\end{equation}
and
\begin{equation}
\vec \eta \equiv \begin{pmatrix}x\\ s\\ v\\ \end{pmatrix} \ .
\end{equation}

Now, introducing equilibrium coordinates
\begin{eqnarray}
u_1&=&x-x_0\\
u_2&=&s\\
u_3&=&v-v_0 \ , 
\end{eqnarray}
we finally obtain the linearized equations
\begin{eqnarray}
u_1'&=&1\\
u_2'&=&\frac{\gamma_0+\sigma_0-1}{x_0}\,u_2-
\frac{\gamma_0\,\sigma_0}{x_0^2}\,u_3\\
u_3'&=&u_2 \ ,
\end{eqnarray}
with $\sigma_0 = \frac{v_0\,g_{v,0}}{2\,h_0}$.

Notice that the first equation has decoupled from the other two. Hence, it suffices to study the behavior of the subsystem $(u_2,u_3)$. We will use standard results from the theory of dynamical systems to do so (see \cite{strogatz}).

In the vicinity of the fixed point, the behavior of the system can be classified by the eigenvalues of the 
sub-matrix $M_{ij}$, with $i,j=2,3$, with characteristic equation
\begin{equation}
\lambda^2-M_{2s}^{(0)}\,\lambda-M_{2v}^{(0)}=0 \ .
\end{equation}

Since we are only interested in values of $v$ and $w$ (the equation of state parameter) which satisfy  $v>1/2$ and
$-1<w\leq1$ 
(see~\cite{Carroll:2003st,Cline:2003gs,Hsu:2004vr} for arguments why this is a sensible choice, 
and~\cite{Carroll:2004hc,Lue:2004za} for how one may be tricked into inferring values outside this
range when considering gravitational theories other than General Relativity), 
we have that $\gamma_0>1$. Furthermore if we choose $v_0=s_2$ (see Section 
\ref{sec:vacuum}), the
value of the bigger power law, typically
$\sigma_0>0$.  This implies that both $M_{2s}^{(0)}$ and $M_{2v}^{(0)}$ are
negative, so that $s_2$ is either a stable node or a stable spiraling
helix. The latter case is tantamount to stability.

We have a stable spiralizing helix if
\begin{equation}
\label{condi}
(M_{2s}^{(0)})^2+4\,M_{2v}^{(0)} < 0.
\end{equation}
If not, we get a stable node. Relation~(\ref{condi}) leads to the
condition
\begin{equation}
(\gamma_0+\sigma_0-1)^2-4\,\gamma_0\,\sigma_0<0\ ,
\label{spirale}
\end{equation}
which holds for all $x_0\ll1$. 

Applying the last condition to the case $\alpha=1$ and
$w=0$ (dust), we have that (see Appendix \ref{app1})
\begin{equation}
\sigma_0 = \frac43\,s_2\,(s_2-0.5)=16.25
\end{equation}
and 
\begin{equation}
\gamma_0 = 3\,s_2+4=15.25.
\end{equation}
For these values of $\sigma$ and $\gamma$, (\ref{spirale}) is easily satisfied, 
so that the power law solution $s_2=3.75$ is a stable spiraling helix.

\end{document}